# Unfolding Large Biomolecules

Erik W. Streed[1,2]

[1]Centre for Quantum Dynamics, Griffith University, Brisbane QLD, 4111 Australia
[2]Institute for Glycomics, Griffith University, Gold Coast QLD, 4222 Australia

*Abstract Summary (35 words)*

*The conformational dynamics of biomolecules drives the chemistry of life. We propose trapping large biomolecular ions in a Paul trap to probe their dynamics and that of their surrounding solvent cage.*

## I. Introduction

The functionality of biological molecules such as nucleic acids, proteins, and carbohydrates are driven by both their chemical composition and conformation. We propose levitating single large isolated biomolecules in an ion trap and using this uniquely adaptable gaseous phase environment to optically investigate their higher-order structure. This approach combines techniques from mass spectroscopy for biomolecule manipulation with single molecule fluorescence imaging approaches for structure detection. Increasing the net charge on the ion will also foster denaturing through Coulomb repulsion. Adding and removing single charges can thus measure the energetics and reversibility of folding processes. Environmental conditions ranging from gas-phase to fully solvated can be probed by varying the number of water molecules adhered to the biomolecular ion. Varying the temperature gives information regarding the entropy of the molecular states. For ions in the gas-phase or not fully solvated the lower end of the workable temperature range is extended beyond the aqueous freezing point.

Ion trapping is a well-established technique for mass spectroscopy and high precision quantum measurements. Charged particles ranging from single electrons [1] to macroscopic dust particles [2] can be trapped for up to months at a time. Mass spectroscopy of large biomolecular ions relies on the capability to charge the particles without disruption of their delicate internal structure. Electrospray ionization (ESI) and matrix-assisted laser desorption/ionization (MALDI) are two commonly used soft ionization techniques that are used to load large, complex biomolecules such as double stranded DNA [3] or whole virus particles [4] into charged particle mass spectrometers. Above the micron level, macroscopic particles such as lycopodium pollen spores [5] can be directly ionized without the use of a solid (MALDI) or liquid (ESI) host medium. Friction with the walls of a plastic syringe when being puffed into a trap provides sufficient tribo-electric charging to remove tens of thousands of electrons per particle [2,5].

Advances in trapped atomic ion fluorescence imaging have recently demonstrated high collection efficiency [6] and wavelength scale resolution [7] with large numerical aperture optics. Existing single molecule and super-resolution fluorescence techniques from wet condensed in-vitro type environments can thus be adapted to optically monitor changes in the conformation within the ion trap environment. The value of the charge-to-mass ratio can also be measured with high precision through detection of the ion's micromotion or via resonant parametric excitation. For larger biomolecules with an appreciable optical scattering cross-section, absorption imaging is also an intriguing option to detect conformal changes. DNA folded into chromosomes is host to numerous additional atomic ions and organic molecules that impact the structure's stability. Ref [8] calculated that absorption imaging at 260 nm should be capable of detecting the unwrapping of 100 nm and 30 nm DNA nucleosomes. Within an ion-trapping environment the additional molecules involved in the folding stablisation process could either be captured separately during the unwrapping process or the sequence of their mass loss quantitatively measured. In addition the lack of surrounding solution minimizes the complications from absorption and scattering.

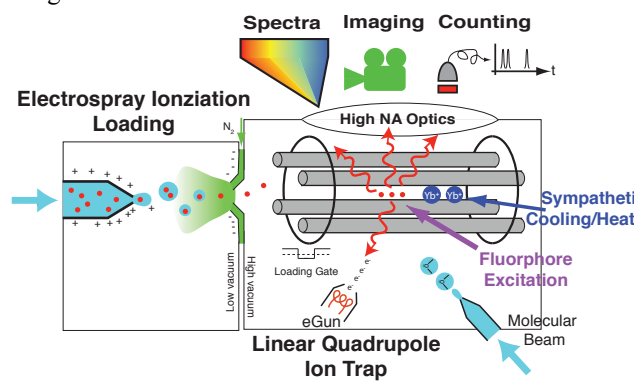

Figure 1. Biomolecular ion loading and trapping apparatus. Biomolecules (red) in solution (blue) are electrospray ionised in a low vacuum chamber against a counter-propagating stream of dry nitrogen (green). The ions are focused through a differential pumping aperture for transient loading into a linear quadrupole ion trap with a time gated voltage inversion on the end-ring electrode. Ions are manipulated through interaction with molecular beams, electron scattering, and laser excitation. Fluorescence collected with high Numerical Aperture (NA) optics can be analysed in the time, wavelength, or spatial domains. Sympathetic thermalisation with trapped atomic ions (e.g. $Yb^+$ or $Ba^+$) can provide indirect laser cooling or heating in a high vacuum environment without the complication of varying a buffer gas temperature.

Fig. 1 depicts an apparatus for investigation of biomolecular folding in an ion trap. The four key components of this device are the mechanisms for loading, confinement, manipulation, and detection of the ions. The mechanics of loading and confining biomolecular ions is well-established technology and will not be covered further. Approaches to manipulation of the biomolecular ion's internal states and fluorescence detection are discussed in subsequent sections.

## II. Manipulation of Biomolecular Ions

Changing the shape of a biomolecule allows us to investigate the processes which govern its structure and dynamics. X-ray crystallography can provide a detailed information about the electron distribution of a crystalised molecule, but it does so for a static configuration and lacks information about the dynamic flexing and other conformational changes the molecule can undergo. These are particularly important when a conformation change is associated with a biologically relevant process such as drug binding or an enzymatic catalysis. The transformation from physiologically active native state to a detectably denatured variant depends on the mechanisms available to precipitate such a change. In solution this change is accomplished by varying parameters including pH, temperature, solvent composition, and concentration of other ion species with response measured through solubility, optical or NMR spectra, optical activity, or reactivity. The ensemble nature of these measurements can make it challenging to interpret the results in the context of biomolecular structure, such as salting-in and salting-out curves for solubility. In mass spectroscopy the fragmentation pattern from the disruption of covalent bonds can give clues to the higher order structure, however it is dependent on the electron rearrangement behavior during the fragmentation process.

With a single biomolecular ion confined in a trap, the charge state can be changed a single electron at a time. The Coulomb force from the distribution of the net charge competes with the internal forces of interest binding the molecule in a particular set of conformations. Fig. 2 illustrates how the hierarchy of structures would be ideally unraveled through this Coulombic mechanis. In the large charge and high temperature limit mass spectroscopy style fragmentation of the biomolecular ion into multiple species occurs at a rate related to the amount of Coulomb repulsion, available thermal energy, and relative favourability of the fragmented structure. Since the net charge of the molecule is only changed in single electron steps, this provides a known self-calibration for the effect of the change in charge. Electrons can be removed selectively through photo-ionisation or less discriminately with low energy electron scattering collisions. Deliberate driving of the ion's motion in a buffer gas filled trap environment has been observed to increase the charge through the tribo-electric effect [2], while an ion near rest will have its net charge quenched through buffer gas collisions. These later processes offer a much gentler alternative to charging the ion through the use of collisional friction interactions.

The success of this process depends highly on the availability to add or remove charge without adverse effects to the overall structure. If the change in charge comes from disrupting a covalent bond, the primary structure will be altered. Unlike in aqueous solution, the difference between Lewis and Brøstend-Lowry acids and bases has an important distinction. In trapping Lewis type systems the donation or acceptance of additional electrons is easily reversible. The donation or acceptance of additional protons (H+) in a Brøstend-Lowry type dynamic may not be either possible (acceptance) or reversible (donation) due to loss when a proton is liberated, either as a charged particle or a neutral. This is not surprising given that multiple protonation states are readily seen as a side effect of electrospray loading [9].

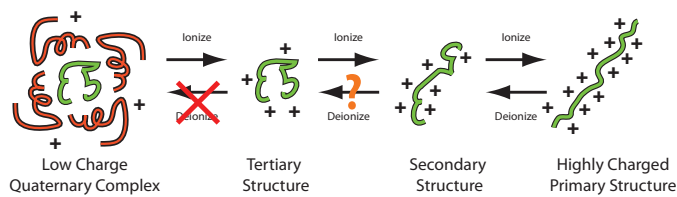

Figure 2. Charge denaturing via Coulomb repulsion. Higher-order structure is unraveled as the Coulombic energy exceeds the binding energies.

No biomolecule functions in isolation. Understanding the impact of the surrounding environment is thus a crucial aspect to unraveling its dynamics. Water molecules prefer proximity to the surface of a highly soluble biomolecule rather than bulk water. This affinity results in the biomolecule moving within a loose "cage" of solvent layers surrounding it. A particularly dramatic instance of the importance of solvation is found in green fluorescent protein. While the fluorophore dipole is buried deep inside this protein, it does not exhibit fluorescence outside of solution. As mentioned above, aqueous solution also provides a reservoir of protons and other species that influence the internal equilibria.

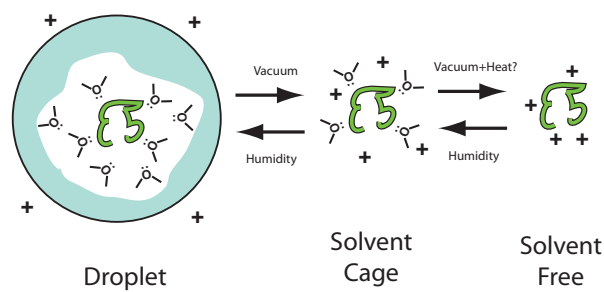

Figure 3. Solvent cage manipulation. Exchange of water molecules with the background gas can either precipitate a droplet or render a dehydrated ion.

Within an ion trap environment the level of solvation can be manipulated by varying the partial pressure of water vapour above or below saturation. Fig. 3 illustrates the inter-conversion stages between a solvated droplet and a solvent free biomolecule. Higher humidity will promote condensation of water molecules on the biomolecular ion, leading to the same mechanics that are observed in droplet formation in the highly solvated limit. Reducing the available water vapour will promote evaporation. The more tightly bound final layers of water in the solvent cage may require an extended duration to evaporate off unless additional energy is added to liberate them. Desorption could also be selectively encouraged by illuminating the molecule/solvent system with infrared lasers in the O-H stretching water absorption bands at 1.95 or 2.9 μm. The later wavelength was successfully used for IR laser based MALDI from water ice [10], a situation similar to the semi-crystallised solvent cage state. Complicating matters the presence of water layers around the biomolecule will likely act as a shield against charge manipulation techniques. Conversely the presence of water provides a substrate acid/base chemistry.

## III. Fluorescence Detection

Single molecule fluorescence techniques can provide a sensitive and non-destructive measure of the internal dynamics of large biomolecules within an ion trap. Single fluorophore systems are easier to realise [9] but have demanding absolute localisation requirements to be suitable for super-resolution imaging [10] that may not be feasible for trapped biomolecular ion experiments. Wavelength-scale resolution trapped atomic ions have realised fluorescence imaging with one nanometer accuracy (FIONA), however this was achieved at a temperature of a few milli-Kelvin (fitting of Fig 2d from [8]). Biomolecular systems with multiple fluorophores are less ubiquitous, but greatly simplify the technical demands on ion trapping and detection. Förster Resonance Energy Transfer (FRET) enables precision proximity measurements at the 4-7 nm range [11] for fluorophore pairs with an overlap in their emission and excitation spectra. Greater distances between fluorophore pairs with spectrally separable emission can be resolved through colocalisation techniques at the tens or hundreds of nanometer level.

FRET provides a rapid measure of the distance between fluorophores in the ratio of their relative emission rates. When the two fluorophores are sufficiently close emissions from the higher wavelength "donor" fluorophore are absorbed and re-emitted by the lower wavelength "acceptor" fluorophore. The effective detection range occurs when the donor only fractionally converts into an acceptor photon. The dipole emission and absorption patterns also make FRET to changes in orientation of the fluorophores. This makes FRET highly sensitive, though complicated in precisely disentangling distance from orientation effects without additional polarisation information. An advantage of this approach is that imaging isn't required, only collection and spectral separation of the fluorescence. This makes the common-mode motion of the biomolecular ion unimportant to detection, so long as the overall excursion does not exceed the collection area of the detector. The time domain variations in the emission ratio provide information on the roto-vibrational modes linking the two groups.

Colocalisation is a more flexible but less sensitive super-resolution imaging method of measuring distances between fluorophore pairs. For each fluorophore an exposure with multiple photons is acquired and a centroid location calculated. The technique relies on imaging with a sufficient degree of chromatic correction, to prevent aberrations such as lateral colour from distorting the signals. The sensitivity to separation measurement depends on the imaging system resolution and the number of photons that can contribute to estimating the centroid location. The relative distances between the fluorophore locations can then be compared. Motion of the ion below the optical resolution limit does not appreciably degrade the centroid accuracy so long as the resolution is the dominant contribution to the RMS width of the imaged spot. The ion motion blurring and the resolution limit are both averaged away in finding the centroid. This is unlike deterministic absolute positioning super-resolution techniques such as STimulated Emission Depletion (STED) or Ground State Depletion (GSD), in which the motion of the ion must be reduced below the position sensitivity. Colocalisation provides a projection of the relative fluorphore positions onto a particular plane. An advantage of the ion trap environment over condensed phase experiments is that the orientation of the biomolecular ion can be controlled by varying the trap parameters, thus facilitating recovery of the three dimensional positioning in colocation imaging of a trapped biomolecule. A biomolecule with a permanent electric dipole will orientate itself along the direction of the residual uncompensated electric field. In the case of no electric dipole or the residual electric compensated to zero the direction of the strongest quadrupole moment will likewise align itself with the weakest axis of confinement, and conversely the weakest quadrupole moment will align itself with the strongest axis of confinement.

Selecting biomolecules for preliminary experiments attempts to balance their viability to demonstrate an effect with utility of that effect. Large biomolecules can fluoresce due to either the composition of their constitute components (ex. Phenylalanine, Tyrosine, and Tryptophan amino acids in proteins), the nature of their final functional form (green fluorescent protein), or through the intentional addition of dye groups (PicoGreen™ in DNA). Intrinsic fluorescence is preferable due to the lack of impact on the structure, however the fluorescence dipole may not be located in an interesting location spatially or spectrally and the likelihood of a pair of fluorescent groups suitable for FRET is low.

## Acknowledgment

Funded by Griffith University Strategic Investment in the Physical Sciences. The author thanks John Chodera for his helpful discussions and introductions into the protein folding community.

## References

[1] G. Gabrielse, H. Dehmelt, and W. Kells, "Observation of a Relativistic, Bistable Hysteresis in the Cyclotron Motion of a Single Electron" *Phys. Rev. Lett.* **54**, 537 (1985).

[2] W. Winter & H. W. Ortjohann "Simple demonstration of storing macroscopic particles in a 'Paul trap'" *Am. J. Phys.* **59** 807 (1991)

[3] F Kirpekar, S. Berkenkamp, F Hillenkamp, "Detection of Double-Stranded DNA by IR- and UV-MALDI Mass Spectroscopy" *Anal. Chem.* **71**, 2334 (1999)

[4] M. A. Tito, K. Tars, K Valegard, J. Hajdu, C. V. Robinson "Electrospray Time-of-Flight Mass Spectrometry of the Intact MS2 Virus Capsid" *J. Am. Chem. Soc.* **122**, 3550-3551 (2000)

[5] M. Kumph, M. Brownnutt, R. Blatt "Two-Dimensional Arrays of RF Ion Traps with Addressable Interactions" *New J. Phys.* **13**, 073043 (2011)

[6] E. W. Streed, B. G. Norton, A. Jechow, T. J. Winehold, D. Kielpinski "Imaging of Trapped Ions with a Microfabricated Optic for Quantum Information Processing" *Phys. Rev. Lett.* **106**, 010502 (2011)

[7] A. Jechow, E. W. Streed , B. G. Norton, M. J. Petrasiunas , & D. Kielpinski, "Wavelength-scale imaging of trapped ions using a phase Fresnel lens" *Opt. Lett.* **36,** 1371 (2011)

[8] E. W. Streed, A. Jechow, B. G. Norton, D. Kielpinski "Absorption imaging of a single atom" *Nat. Comm.* **3** 933 (2012)

[9] D. Offenberg, C. B. Zhang, Ch. Wellers, B. Roth, S. Schiller "Translational cooling and storage of protonated proteins in an ion trap at subkelvin temperatures" *Phys. Rev. A* **78**, 061401R (2008).

[10] A. Pirkl, J. Soltwisch, F. Draude, K. Dreisewerd "Infrared matrix-assisted laser desorption/ionization orthogonal-time-of-flight mass spectrometry employing a cooling stage and water ice as a matrix." *Anal. Chem.* **84** 5669 (2012).

[11] H. E. Grecco, P. J. Verveer "FRET in Cell Biology: Still Shining in the Age of Super-Resolution?" *Chem. Phys. Chem*. **12** 484 (2011)